\documentclass[aps,prd,twocolumn,letter]{revtex4}
\usepackage{graphicx}
\def\physrep{Phys. Rep.}
\def\mnras{MNRAS}

\newcommand{\eqref}[1] {equation $($\ref{#1}$)$}
\def\beq{\begin{equation}}
\def\eeq{\end{equation}}
\def\beqn{\begin{eqnarray}}
\def\eeqn{\end{eqnarray}}

\def\cl{C_{\ell}}

\def\O{\Omega}

\def\Om{\ensuremath{\Omega_{\mathrm{m}}}}
\def\Ob{\ensuremath{\Omega_{\mathrm{b}}}}
\def\Oc{\ensuremath{\Omega_{\mathrm{CDM}}}}

\def\2gcm{\textrm{g cm$^{-2}$}}

\def\av#1{\l \langle{#1}\r \rangle}
\def\l{\left}

\def\r{\right}

\def\Sig{\Sigma}

\def\H0{\ensuremath{\mathrm{H}_0}}
\def\nhat{\hat{\bmm{n}}}

\def\nn{\nonumber}

\def\fsky{f_{\mathrm{sky}}}

\newcommand{\bmm}[1]{{\mathbf{#1}}}

\def\gcmb{\ensuremath{g_{\mathrm{\scriptscriptstyle{CMB}}}}}
\def\ggal{\ensuremath{g_{\mathrm{\scriptstyle gal}}}}
\def\kcmb{\ensuremath{\kappa_{\mathrm{\scriptscriptstyle{{CMB}}}}}}
\def\kgal{\ensuremath{\kappa_{\mathrm{\scriptscriptstyle{gal}}}}}

\def\nl{N_\ell}
\def\pc{\%}
\def\ode{\Omega_{d}^{e}}

\begin{document}
\title{Measuring Distance Ratios with CMB-Galaxy Lensing Cross-correlations}
\author{Sudeep Das}
\email{sudeep@astro.princeton.edu}
\author{David Spergel}
\email{dns@astro.princeton.edu}
\affiliation{Princeton University Observatory\\
Peyton Hall, Ivy Lane, Princeton, NJ 08544 USA}
\begin{abstract}
{We propose a  method for cosmographic measurements by combining gravitational lensing of the cosmic microwave background (CMB) with cosmic shear surveys. We cross-correlate the galaxy counts in the lens plane with two different source planes: the CMB at $z \sim 1100$ and galaxies at an intermediate redshift. The ratio of the galaxy count/CMB lensing cross-correlation to the galaxy count/galaxy lensing cross correlation is shown to be a purely geometric quantity, depending only on the distribution function of the source galaxies. By combining Planck, ADEPT and LSST the ratio can be measured to $\sim 4\%$ accuracy, whereas a future polarization based experiment like CMBPOL can make a more precise ($\sim 1\%$) measurement. For cosmological models where the curvature  and  the equation of state parameter are allowed to vary, the  direction of degeneracy defined by the measurement of this ratio is different from that traced out by Baryon Acoustic Oscillation (BAO) measurements.  Combining this method with  the stacked cluster mass reconstruction cosmography technique as proposed by Hu, Holz and Vale (2007), the uncertainty in the ratio can be further reduced, improving the constraints on cosmological parameters. We also study the implications of the lensing-ratio measurement for early dark energy models, in context of the parametrization proposed by Doran and Robbers (2006). For models which are degenerate with respect to the CMB, we find both BAO and lensing-ratio measurements to be insensitive to the early component of the dark energy density. }
\end{abstract}
\date{\today}
\maketitle
\section{Introduction}
Weak gravitational lensing of the cosmic microwave background (CMB) (see \cite{2006PhR...429....1L} for a review) provides us with a unique opportunity to study the large scale distribution of dark matter in the universe out to much greater distances than accessible through conventional galaxy-lensing studies. It has been shown \citep{2003PhRvL..91n1302J} that by studying the gravitational lensing of galaxies in different redshift slices by the same foreground structures, the geometry of the universe and eventually, dark energy evolution may be constrained --- a method known as cross-correlation cosmography. In this paper, we propose a similar method in which we treat the CMB as one of the background slices. This not only provides an extremely well standardized distance to compare other distances to, but also incorporates the longest possible distance in the ratio of distances probed by this method, making it a more sensitive probe of cosmological parameters than ratios involving distances restricted to galaxy surveys.  Recently, Hu et al.  \citep{2007PhRvD..76l7301H} have proposed a method for measuring the same lensing-ratio  by comparing the convergence profile of the a cluster reconstructed via background galaxy shear with that reconstructed via CMB lensing, and then stacking several clusters to improve the precision of the measurement. The method we propose here depend on cross-correlations rather than reconstruction of convergence of individual objects, and will have different systematics. As such, this method is a powerful complement to the cluster-lensing based method and the two may be combined to obtain more precise measurements of the ratio. 
\section{Lensing Ratio: The key observable}
Cosmological weak lensing effects are conveniently encoded in the effective convergence field, which is defined as a weighted projection of the matter overdensities $\delta$ \citep{2001PhR...340..291B},
\beq
\kappa(\nhat)=\frac32 \Om H_{0}^{2}\int  d\eta d_{A}^{2} (\eta) \frac{g(\eta)}{a(\eta)}\delta(d_A(\eta) \nhat,\eta),
\eeq 
with
\beq
g(\eta)=\frac{1}{d_A(\eta)}\int_{\eta}^{\infty}d\eta' W_b(\eta') \frac{d_A(\eta'-\eta)}{d_A(\eta') }
\eeq
where $d_A(\eta)$ is the comoving angular diameter distance corresponding to the comoving distance $\eta$. Here,  $a(\eta)$ is the scale factor, while $\Om$  and $H_{0}$ represent the present values of the matter density parameter and the Hubble parameter, respectively.
The quantity $g(\eta)$ represents the fact that  sources are distributed in comoving distance with a normalized distribution function $W_b$. Since the CMB photons all come from nearly the same cosmological distance, we can approximate the source distribution function as, $W_{b}(\eta) \simeq \delta(\eta - \eta_{0})$, giving 
\beq
\gcmb(\eta)=\frac{d_A(\eta_0-\eta)}{d_A(\eta_0) d_A(\eta) },
\eeq
where $\eta_{0}$ is the comoving distance to the last scattering surface.
We will denote the same quantity for a background galaxy population with redshift distribution $p_g(z)dz=W_b(\eta)d\eta$,  with the symbol $\ggal(\eta)$. \par
We also consider a suitable foreground population as a tracer of large-scale structure. The projected fractional overdensity of the tracers can be written as,
\beq
\Sigma(\nhat)=\int d\eta  W_f(\eta) \delta_g(\eta \nhat,\eta),
\eeq  
where $\delta_g$  represents the fractional tracer overdensity  and $W_f$ is the normalized tracer distribution function in comoving distance. We assume that the Fourier modes of the tracer overdensity field are related to those of the underlying matter density field via a scale and redshift dependent bias factor, so that $\delta_{g}(\bmm k,\eta) = b(k,\eta) \delta(\bmm k,\eta)$.
If we cross-correlate the tracer overdensity map with the convergence field, we obtain the cross power spectrum,
\beq
\cl^{\kappa\Sigma}=\frac{3}{2}{\O_m H_{0}^{2}}  \int d\eta b_\ell(\eta) W_f(\eta) \frac{g(\eta)}{a(\eta)}  P(\frac{\ell}{d_A},\eta),
\eeq
where we have used the Limber approximation and the orthogonality of spherical harmonics. We have also introduced the shorthand notation, $b_\ell(\eta) \equiv  b(\frac{\ell}{d_A},\eta)$.\par
Now, we will introduce two separate cross-correlation measures involving the foreground tracer population. First, we consider the case for the CMB as the  background source. By constructing estimators out of quadratic combinations of  CMB fields (temperature and polarization), it possible to obtain a noisy reconstruction of the convergence field (note that the actual observable in this case is the deflection field) out to the last scattering surface \citep{2002ApJ...574..566H,2003PhRvD..67d3001H}, which we denote as $\kcmb$. The  power spectrum of the noise in the reconstruction, $N_\ell^{\kcmb\kcmb}$, can be estimated knowing the specifications for the CMB  experiment. The cross-correlation of the reconstructed convergence field from the lensed CMB with the foreground tracer, gives the signal,
\beq
\cl^{\kcmb\Sigma}=\frac{3}{2}{\O_m H_{0}^{2}}  \int d\eta b_\ell(\eta) W_f(\eta) \frac{\gcmb(\eta)}{a(\eta)}  P(\frac{\ell}{d_A},\eta),
\eeq
where we have used the source distribution kernel $\gcmb$ appropriate for the CMB being the background source. \par
Next, we consider the case for the weak lensing of background galaxies. The relevant observable in this case is the traceless symmetric shear field on the sky, the measurement of which allows a noisy reconstruction of the convergence field appropriate to the background galaxy distribution, $\kgal$.  In this case, the noise is primarily due to intrinsic ellipticity of the background galaxies and has the spectrum, $N_l^{\kgal\kgal}=\av{\gamma^2_{\mathrm{int}}}/\bar{n}$ where ${\av{\gamma^2_\mathrm{int}}}^{1/2}\sim 0.3$ and $\bar{n}$ is the number of background galaxies per steradian \citep{1992ApJ...388..272K}. If we cross correlate this convergence field with the foreground tracers, we find the signal, 
\beq
\cl^{\kgal\Sigma}=\frac{3}{2}{\O_m H_{0}^{2}}  \int d\eta b_\ell(\eta) W_f(\eta) \frac{\ggal(\eta)}{a(\eta)}  P(\frac{\ell}{d_A},\eta),
\eeq
where we have used the source distribution kernel, $\ggal$ appropriate for background galaxies. \par
If the foreground distribution is narrow in redshift so that it can be approximated by a delta function, $W_f(\eta)\simeq \delta(\eta-\eta_f)$, then the ratio of the above two cross-correlation measures, which we call the lensing-ratio, reduces to,
\beq
\label{raw-ratio}
r\equiv\frac{\cl^{\kcmb \Sigma}}{\cl^{\kgal\Sigma}}=\frac{\gcmb(\eta_f)}{\ggal(\eta_f)}
\eeq
which is simply the geometrical ratio of the source distribution kernels.
If the background galaxy distribution, too, is sufficiently narrow in redshift around $z=z_{\mathrm gal}$, this becomes,
\beq
 r= \frac{d_A(\eta_0-\eta_f)d_A(\eta_{\mathrm{gal}})}{d_A(\eta_{\mathrm{gal}}-\eta_f)d_A(\eta_0)}.
\eeq 
Note that this is independent of the angular scale, tracer bias and the power spectrum. Therefore, measurements at several multipoles can be combined to constrain the lensing-ratio. Since the distance ratios depend on the cosmology and specifically on the dark energy model, this can be used to constrain dark energy parameters.
\section{Upcoming surveys and a new  probe of dark energy and curvature}
Large scale structure surveys, together with precision measurements of the CMB anisotropies have already provided  us with a wealth of knowledge about the geometry, evolution and composition of the Universe. In the coming decade, Cosmologists will carry out even larger scale galaxy and lensing surveys and produce higher resolution CMB maps. We consider a combination of three experiments in order to assess how well the lensing-ratio can be measured in such future surveys. We consider the redshift slice of foreground tracers (lenses) to be drawn from an ADEPT-like \footnote{Advanced Dark Energy Physics Telescope; http://universe.nasa.gov/program/probes/adept.html} large scale structure survey and the background (source) galaxies taken from an LSST-like \footnote{Large Synoptic Survey Telescope; http://www.lsst.org}  weak lensing experiment. For the CMB lensing measurements, we consider the upcoming Planck mission as well as a prospective polarization-based mission like CMBPOL.\par
\begin{figure}[htbp]
\begin{center}
\includegraphics[scale=0.39]{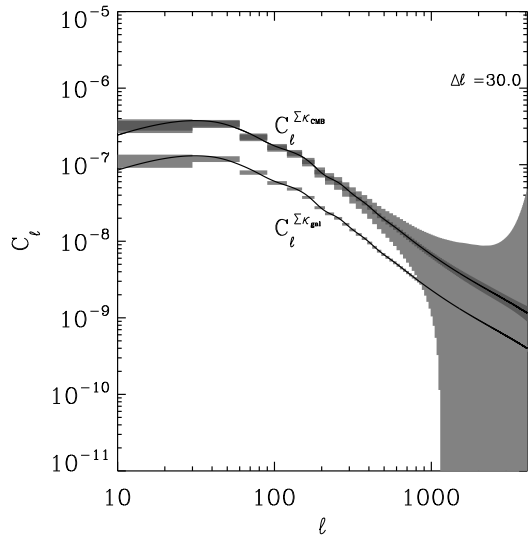}
\caption{Cross power spectra, the ratio of which is being studied (cf. equation~\ref{raw-ratio}). Also shown are predicted $1\sigma$ errors in uniform bins of size $\Delta \ell = 30$. For the $\cl^{\Sigma\kcmb}$ case, the outer (lighter) errors correspond to lensing reconstruction using temperature and polarization with Planck, while the inner (darker) ones correspond to the same for CMBPOL.\label{twoCls}}
\end{center}
\end{figure}
The foreground galaxy slice is taken as a step function in the redshift range $(0.8,0.9)$ with $350$ galaxies per square degree. The source galaxies are also assumed to be distributed uniformly in redshift, between $z=1.2$ and $1.6$ with a number density of $40$ galaxies per square arcmin. We model Planck to be a $7^{\prime}$ FWHM instrument with temperature and polarization sensitivities of $28$ and $57$ $\mu$K-arcmin, respectively. For CMBPOL, we adopt a $3^{\prime}$ beam FWHM and temperature and polarization sensitivities of $1$ and $1.4$ $\mu$K-arcmin, respectively. We assume that both CMB experiments cover $65\%$ of the sky and all cross-correlations are performed over the same area. For calculations performed here we assumed a WMAP 5-year normalized $\Lambda$CDM cosmology with $\Ob h^{2}=0.0227$, $\Oc h^{2}=0.1099$, $\Omega_{\Lambda}=0.742$, $\tau=0.087$, $n_{s} = 0.963$ and $A_{s}=2.41\times 10^{-9}$.\par
In Fig.~\ref{twoCls}, we display the two cross power spectra appearing in the defining \eqref{raw-ratio} of the lensing ratio along with binned uncertainties predicted from the experimental specifications.
\par
\begin{figure}[htbp]
\begin{center}
\includegraphics[scale=0.36]{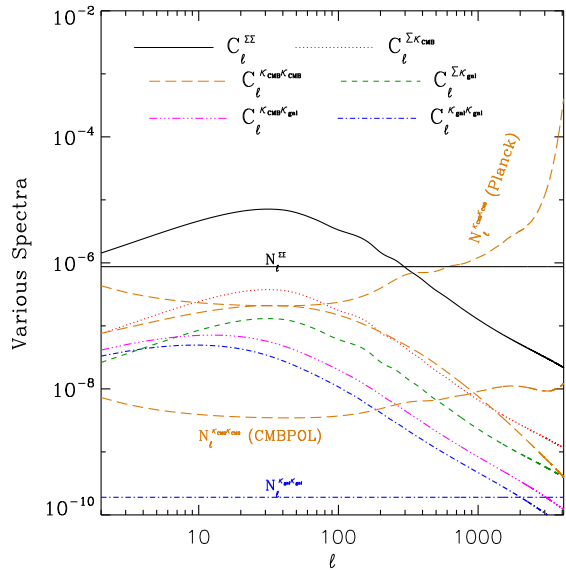}
\caption{Various power spectra that enter the calculation of the error on the lensing ratio (cf. eq.~\ref{ratioerror}). Each of the noise power spectra has been plotted with the same line style as its corresponding signal power spectrum and labeled as $N_\ell$. The noise spectrum for the CMB lensing reconstruction has been indicated both for Planck and CMBPOL.   \label{VariousCls}}
\end{center}
\end{figure}

 The error on the ratio can be obtained as follows. We begin by defining the log-likelihood, 
\beq
\chi^2(r)=\sum_\ell \frac{Z_\ell^2}{\sigma^2(Z_\ell)}
\eeq
where, $Z_\ell=\cl^{\kcmb\Sigma} -r\cl^{\kappa_{\mathrm{gal}} \Sigma}$. We compute the variance of $Z_l$ at the value $r_0$ of $r$ computed in the fiducial cosmology,
\beqn
\label{sigmasquare}
\nn\sigma^2(Z_\ell)&=&\frac{1}{(2\ell+1)\fsky}\l[\tilde \cl^{\kappa_\mathrm{CMB}\kappa_\mathrm{CMB}}\tilde\cl^{\Sigma\Sigma}+(\cl^{\kappa_\mathrm{CMB}\Sigma})^2\r.\nn\\&& +  \l. r_o^2 
\l(\tilde \cl^{\kappa_\mathrm{gal}\kappa_\mathrm{gal}}\tilde\cl^{\Sigma\Sigma}+ (\cl^{\kappa_\mathrm{gal}\Sigma})^2\r)\r. \nn
\\&&\l. -2r_0\l(\cl^{ \kappa_\mathrm{CMB}\kappa_\mathrm{gal}}\tilde\cl^{\Sigma\Sigma}+\cl^{\kappa_\mathrm{CMB}\Sigma}\cl^{\kappa_\mathrm{gal}\Sigma}\r)\r]
\eeqn
where,
$$\tilde\cl^{XX}=\cl^{XX}+N_{\ell}^{XX}$$
include the noise power spectra. The Poisson noise for the foreground tracer is taken as $\nl^{\Sig\Sig} = 1/\bar{n}_{f}$.
Then maximum likelihood estimate for the ratio is then obtained by solving $\partial \chi^{2}(r)/\partial{r} = 0$ to be,
\beq
\hat r = \frac{\sum_{\ell}\cl^{\kappa_\mathrm{CMB}\Sigma}\cl^{\kappa_\mathrm{gal}\Sigma}/\sigma^2(Z_\ell)}{\sum_{\ell} (\cl^{\kappa_\mathrm{gal}\Sigma})^{2}/\sigma^2(Z_\ell)}
\eeq 
Now, we can estimate the error on $r$ as, 
\beq
\label{ratioerror}
\frac{1}{\sigma^2(\hat r)}=\frac12 \frac{\partial^2{\chi^2(r)}}{\partial r^2}=\sum_\ell\frac{(\cl^{\kappa_\mathrm{gal}\Sigma})^2}{\sigma^2(Z_\ell)}.
\eeq
 Various auto, cross and noise power spectra that enter the calculation of the error on $r$ are shown in Fig.~\ref{VariousCls}. The above figures borne out the expected feature that the noise power spectrum in the lensing reconstruction is the largest source of uncertainty that propagates into the error on $r$. 
\par
 The estimated errors on $r$ are shown in Table~\ref{results}. For Planck, we find that the lensing-ratio can be estimated to $\sim 4\%$ while with CMBPOL a $\sim 1\%$ measurement is possible.
\begin{table}[!t]
  \center
  \begin{tabular}{@{\extracolsep{\fill}}cccc}
    Experiment & Type & (S/N)$^{\mathrm{cross}}$ & $\Delta r/r (\%)$ \\
    \hline
    \hline
    Planck & POL & 25.8 & 3.8  \\
           & TT & 23.3 & 4.2 \\
   	CMBPOL & POL & 102.6& 1.0 \\
           & TT &84.5 & 1.2\\
    \hline
    \hline
  \end{tabular}
\caption{\label{results} Predictions for the cross-correlation studies described in the text with foreground galaxies from ADEPT, background galaxies from LSST and different CMB experiments. The quantity (S/N)$^{\mathrm{cross}}$ represents the signal-to-noise ratio in the estimation of the cross correlation between the foreground tracer density with CMB lensing. The last column shows percentage error in the lensing-ratio estimator, r of \eqref{raw-ratio}. We show the prediction for both temperature based (TT) and polarization based (POL) reconstruction of the deflection field from the lensed CMB. } 
\end{table}
\begin{figure*}[t]
\includegraphics[scale=0.77]{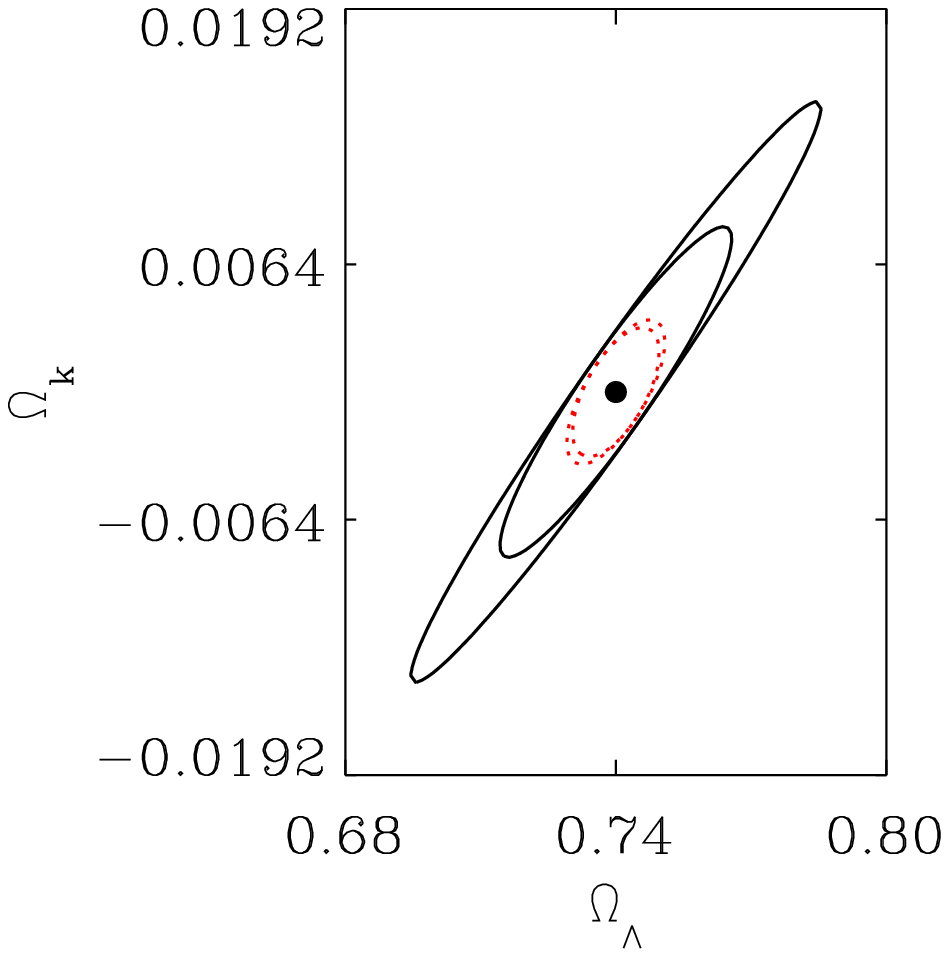}
\includegraphics[scale=0.77]{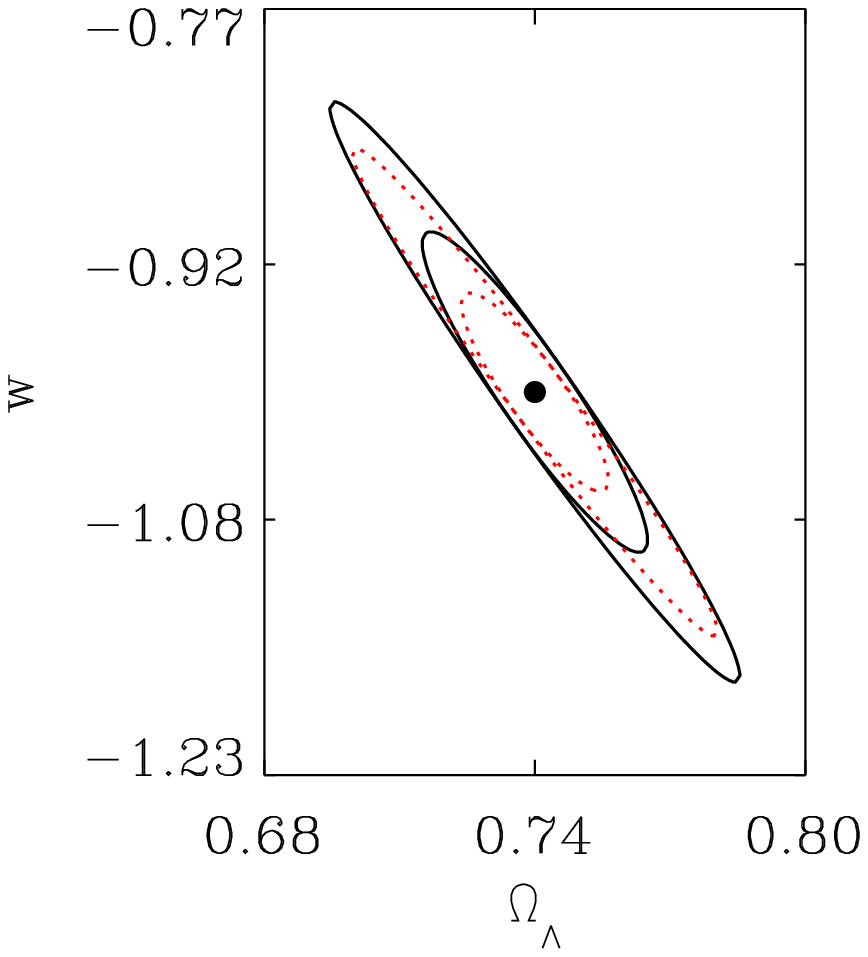}
\caption{\label{wkmodels}\emph{Left Panel}: Improvements of constrains in the $\O_{k}-\Omega_{\Lambda}$ plane for a vacuum energy model with Planck by adding a $1\pc$ measurement of the lensing-ratio. The outer solid contour is the $68\pc$ confidence interval from primary CMB alone while the inner solid contour is the same after adding the lensing-ratio. The dotted contours have the same interpretation but represent the case where information from lensing extraction has been added to the CMB Fisher matrix. \emph{Right Panel:} Same as left, but for the $w-\Omega_{\Lambda}$ plane, assuming flatness. }
\end{figure*}
\section{Parameter constraints}
 For Planck priors, improvements on cosmological parameter constraints upon adding the lensing-ratio to the primary CMB observables become appreciable when the error on the ratio decreases below $10\%$ \citep{2007PhRvD..76l7301H}. It is interesting to note here that the method for estimating the lensing-ratio proposed by Hu, Holz and Vale (2007) \citep{2007PhRvD..76l7301H}, which relies on cluster mass reconstruction can be further improved with the maximum likelihood based estimator proposed by Yoo and Zaldarriaga \citep{2008arXiv0805.2155Y} and can complement the method proposed here. By combining the two methods for the same redshift slices, it may be possible to reduce the uncertainty in the lensing-ratio to percent or sub-percent levels.  \par

In order to assess how a percent-level measurement of the ratio will help constrain a set of cosmological parameters $\{p_{i}\}$ in conjunction with the CMB experiments, we define a Fisher matrix for the lensing-ratio, 
\beq
	F_{ij}^{r} = \frac{\partial{\ln r}}{\partial p_{i}} \frac 1{\sigma^{2}(\ln r)} \frac{\partial \ln r}{\partial p_{j}}.
\eeq
and add it to the Fisher matrix from a CMB experiment. The error in a parameter is then estimated from the inverse of the combined Fisher matrix as $\sigma(p_{i}) = \sqrt{[\bmm F^{-1}]_{ii}}$. We consider two variants of the CMB Fisher Matrix, one with only the primordial power spectra and the other with the power spectra involving the weak lensing deflection field extracted from CMB lensing measurements \citep{2006PhRvD..73d5021L,2006JCAP...10..013P}. We do not consider any foreground contamination in any of these. Fig.~\ref{wkmodels} shows the constraints predicted with Planck specifications and a $1\%$ error on the  lensing-ratio, for minimal extensions to the standard 6-parameter model. These constrains are marginalized over all other parameters. The constraints on curvature assuming $w=-1$ and on $w$ assuming flatness, both improve over the primary CMB case after adding in the ratio. For CMB with lensing extraction the improvement on $w$ is still substantial while that on $\Omega_{k}$ is marginal. \par
\begin{figure*}
\includegraphics[scale=0.77]{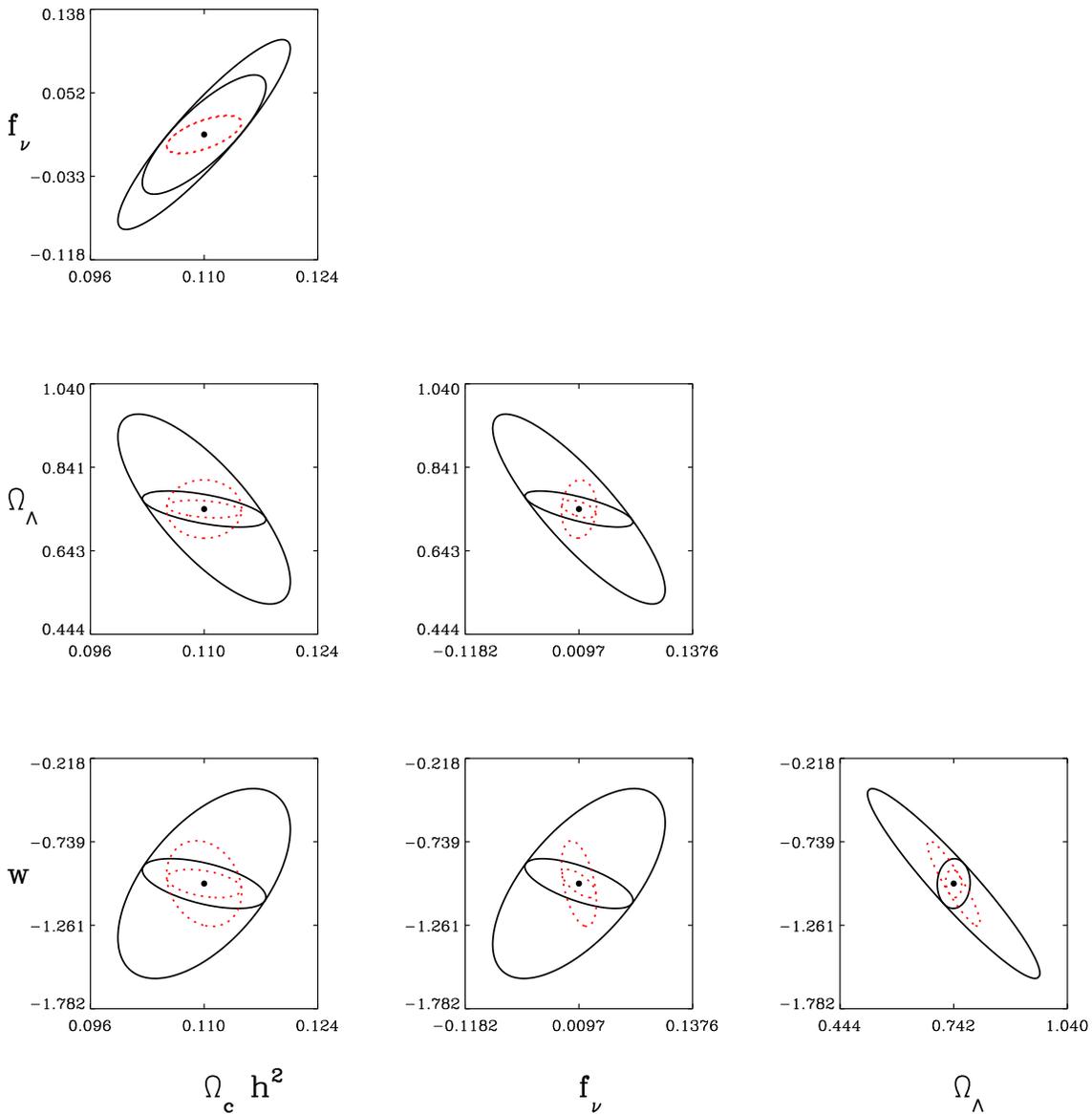}
\caption{\label{wkmnumodels} Improvements in the constraints on the interesting subset of parameters in the eleven parameter model involving massive neutrinos and free dark-energy equation of state (see text). The interpretations of the contours are same as in Fig.~\ref{wkmodels} }
\end{figure*}

Following \cite{2006PhRvD..73d5021L}, we also consider a more general 11-parameter model with $w$, fraction of dark matter in massive neutrinos $f_\nu$, the effective number of neutrino species, the running of the spectral index and the primordial Helium fraction. We adopt a fiducial value of $0.1$ eV for the total neutrino mass. The constraints on the interesting subspace of parameters are shown in Fig.~\ref{wkmnumodels}. 
We find in this case that adding in the ratio significantly improves the constraints on $w$ and $f_\nu$ over the primary CMB case. In fact, for $w$ and $\Omega_\Lambda$ the improvements surpass those from the lensed CMB Fisher matrix. This is particularly interesting because the lensed-CMB-only constraints require an estimate of the convergence field from the four point function in the lensed CMB itself and is more prone to systematics than the cross-correlations that enter the ratio calculation. From Fig.~\ref{wkmnumodels}, it is apparent that  for the CMB Fisher matrix with lensing extraction the constraint on the neutrino mass is rather tight, so that no further gain is obtained by adding in the lensing-ratio. It is important to keep in mind that Fisher matrix methods tend to  overestimate the error on neutrino mass due to non-Gaussianity in the associated likelihood \citep{2006JCAP...10..013P}. Therefore, the errors estimated here are somewhat higher than those predicted from a full Monte-Carlo forecast. \par
\begin{figure}
\includegraphics[scale=0.5]{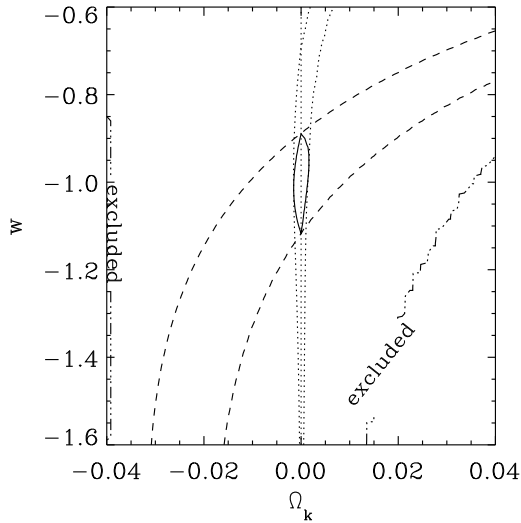}
\caption{\label{baoRatio} Constraints $(68\%)$ in the $w$-$\Omega_{k}$ plane from BAO and lensing-ratio measurements. The dashed line indicate constraints from the lensing-ratio while the dotted line represents the same for the BAO ratio. The solid contour shows the joint constraint. Regions outside the contours labeled ``excluded'' are not allowed due to the strong CMB prior (see text).}
\end{figure}
We next consider models with curvature and a free dark energy parameter $w$. CMB measurements allow for a large degenerate valley in the $w-\Omega_{k}$ plane making Fisher Matrix results deceptive. We treat this model by imposing a strong CMB prior in which we explore the parameter space while keeping the high-redshift variables $\Ob h^{2}$, $\O_{m}h^{2}$ and $\theta_{A}=r_{s}{(z_{0})}/d_{A}(\eta_{0})$ fixed. For each degenerate parameter set, we calculate the value of $r$ as well as the spherically averaged baryon acoustic oscillation (BAO) distance ratio \citep{2007MNRAS.381.1053P,2005ApJ...633..560E}, $r_{s}(z_{d})/D_{v}(z)$, where $r_{s}$ is the comoving sound horizon scale at the drag epoch, $z_{d}$ and $D_{v}$ is an effective distance measure to redshift $z$. We compute it for  $z=1.5$, a typical median redshift for an ADEPT-like survey. The constraints in the $w-\Omega_{k}$ plane from each of these  methods are shown in Fig~\ref{baoRatio} for a $0.4\%$ measurement of the BAO ratio with ADEPT \citep{2007ApJ...665...14S} combined with a $1\%$ measurement of the lensing-ratio. We find that the degeneracy direction for the BAO is quite different from that for the lensing-ratio. Together they can put a $\sim 0.01\%$ limit on $\Omega_{k}$ and a simultaneous $\sim 10\%$ limit on $w$, without adding in any other cosmological prior.\par
Next, we turn to scenarios with early dark energy. In particular, we choose the dark energy parametrization proposed by \citep{2006JCAP...06..026D}, namely, 
\beq
\Omega_{d}(a) = \frac{\Omega_{d}^{0}-\Omega_{d}^{e}(1-a^{-3 w_{0}})}{\Omega_{d}^{0}+\Omega_{m}^{0} a^{3 w_{0}}} + \Omega_{d}^{e} (1-a^{-3w_{0}})
\eeq
where $\Omega_{d}^{0}$ is the present value of the dark energy density function, $\Omega_{d}(a)$, and  $\Omega_{d}^{e}$ is its asymptotic value at high redshift. In this parametrization, the dark energy equation of state $w(a)$ has the value $w_{0}$ at present, crosses over to $w\simeq 0$ during matter domination and goes to $w\simeq 1/3$ in the  radiation dominated era. Two relations of interest in this model that follow from the definition of the Hubble parameter,
\beq
H^{2}(a)=H_{0}^{2}  \frac{(\Om a^{-3} + \O_{r} a^{-4})}{1-\Omega_{d}(a)}
\eeq
are the scaling,  with $\Omega_{d}^{e}$, of the comoving sound horizon at last scattering or the drag epoch, namely,
\beq
\label{soundHorizon}
r_{s}(\Omega_{d}^{e}) = \sqrt{1-\ode} ~~ r_{s}(\ode=0),
\eeq
and the behavior of the comoving angular diameter distance which, for flat cosmologies, given by,
\beq
\label{dAng}
d_{A}(z) = \frac{c}{H_{0}}\int _{0}^{z} dz \frac{\sqrt{1-\Omega_{d}(z})} {[\Om (1+z)^{3}+\O_{r} (1+z)^{4}]^{{1/2}}}. 
\eeq
We consider the parameter space spanned by $(w_{0},\ode)$ and study how well it can be constrained given measurements of the lensing ratio, the CMB and the BAO. We again impose the strong CMB prior, and compute the observables for each point in the $(w_{0},\ode)$ space degenerate with respect to the CMB. Since early dark energy shifts the comoving sound horizon according to \eqref{soundHorizon}, and we are fixing the values $\Om h^{2}$ and $\O_{b} h^{2}$  (which fixes the redshift of last scatter), \eqref{dAng} immediately implies that  the only way to keep the angular scale $\theta_{A}=r_{s}{(z_{0})}/d_{A}(\eta_{0})$ constant is by varying $\Omega_{d}^{0}$ such that $d_{A}$ scales like $\sqrt{1-\ode}$. Thus, quite counter-intuitively, we find that the early value of the dark energy indirectly affects low redshift evolution. An unfortunate consequence of this is that both BAO and lensing ratio measurements are rendered insensitive to the value of $\ode$. As shown by \cite{2008JCAP...06..004L}, it is possible to have a set of cosmological models degenerate with respect to the CMB without having $\Om h^{2}$ strictly constant. For example, holding $\Om$, and therefore $\Omega_{d}^{0}$ constant, it is still possible to have nearly indistinguishable CMB power spectra. In this case, to preserve $\theta_{a}$, one has to change $h$ accordingly. This, again, leads to the shift in low redshift distances and makes our observables insensitive to $\ode$. 
These issues are discussed in some detail in the Appendix.  
\section{Conclusions}
We have proposed a way of measuring a ratio of comoving angular diameter distances that appear in the lensing kernels for CMB and galaxy lensing. By combining Planck, ADEPT and LSST, it is possible make a percent level measurement of this ratio. A polarization based experiment like CMBPOL has the potential of making a more precise measurement. The precision in the measurement can be potentially increased by combining it with cluster mass reconstruction based measurement of the same quantity. The ratio is sensitive to late-time geometry and composition of the Universe and a percent level measurement combined with Planck data can provide interesting constraints and consistency checks, independently of other cosmological probes. By choosing the CMB as one of the lens planes, this method allows higher redshifts to be probed than galaxy-lensing cosmography. 
\par
While the distance ratio is sensitive to late time dark energy, we find it to be  rather insensitive to early dark energy,  particularly for the  parametrization proposed in \cite{2006JCAP...06..026D}. As discussed in the Appendix, when a strong CMB prior is imposed, the  values of low redshift parameters shift in conjunction with the asymptotic high redshift value of early dark energy in such a manner as to render both the BAO ratio and the lensing ratio rather insensitive probes.  This behavior is most likely a specific feature of the said parametrization. The effectiveness of the lensing ratio as a cosmological tool for a wider class of quintessence models remains to be studied. 
\appendix*
\section{BAO and lensing ratios as probes of early dark energy}
\begin{figure*}[htbp]
\begin{center}
\includegraphics[scale=0.39]{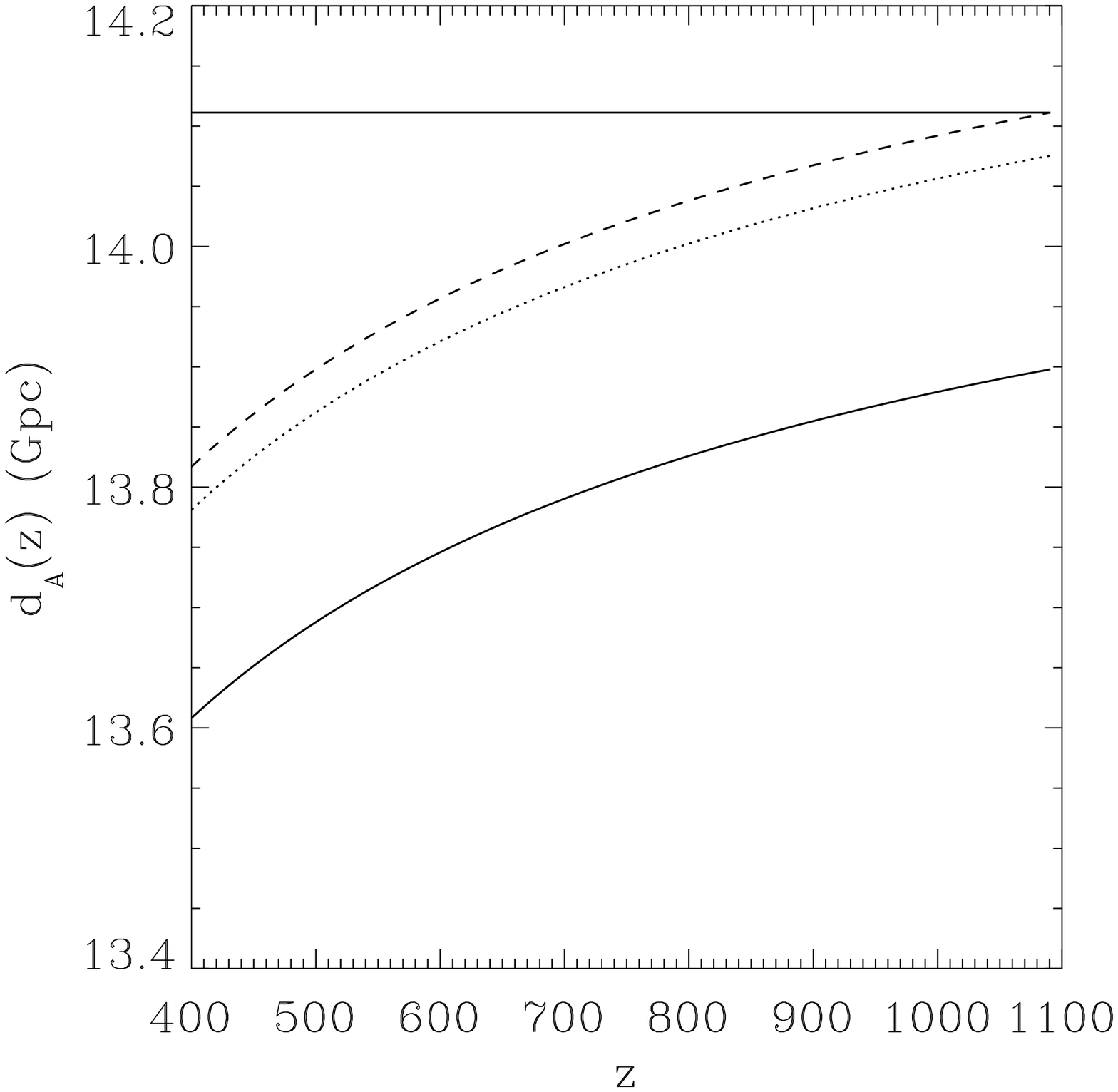}\includegraphics[scale=0.39]{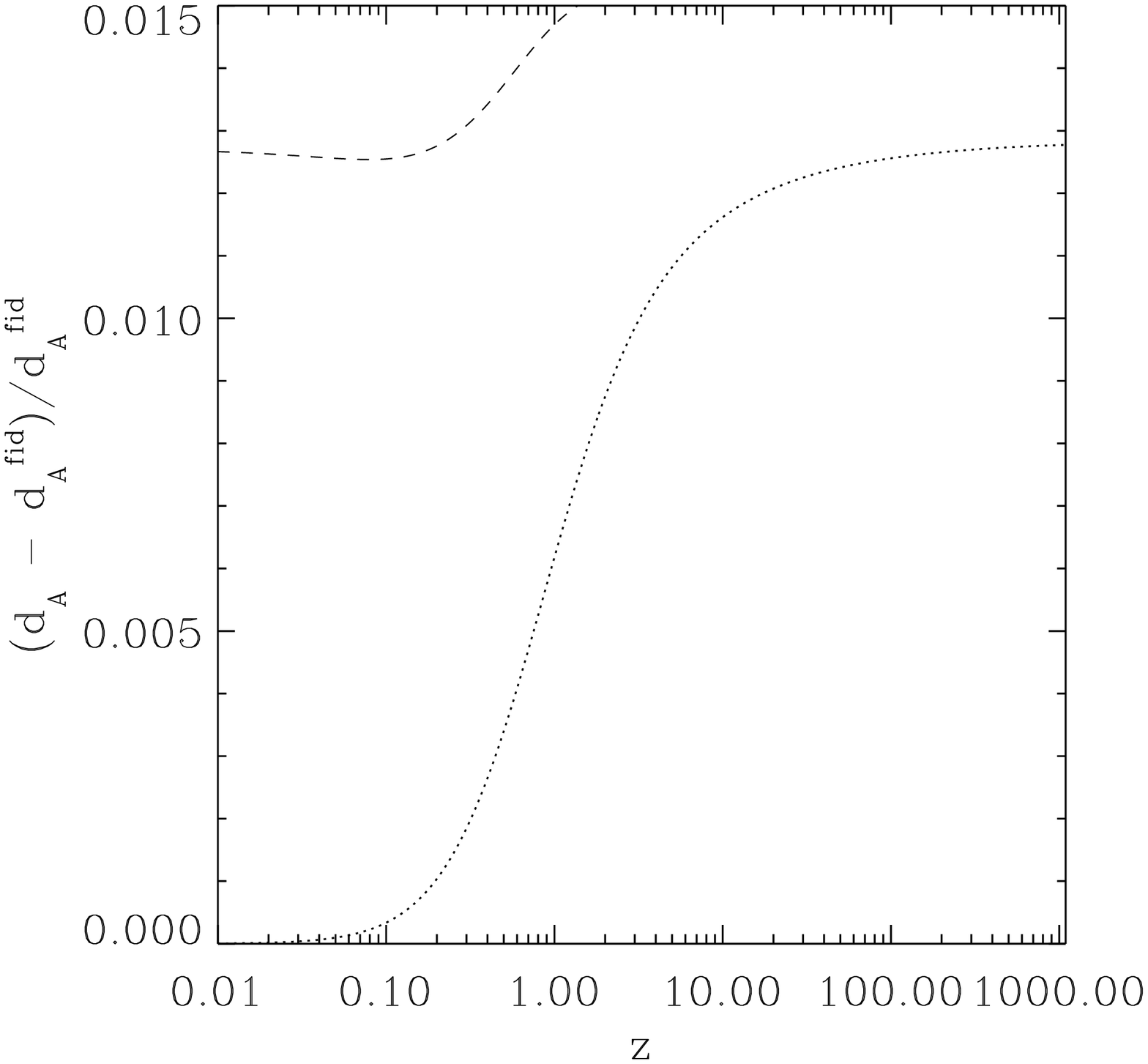}
\caption{{\emph{ Left:}} Comoving angular diameter distance at the high redshift end for various models. The solid curve corresponds to the fiducial early dark energy model with $\O_{d}^{0}=0.742$, $w_{0}=-1$ and $\ode = 0.03$. The horizontal line indicates the value of $d_{A}(z_{0})$ at  the last scattering surface required by a wrong model with $\ode=0$, to keep the CMB angular scale, $\theta_{A}$, constant. The dotted line represents a model with  all parameters kept same as the fiducial model, except $\ode$ which is set to zero. This falls short of the required $d_{A}(z_{0})$ and hence the only free parameter in the model, $\O_{d}^{0}$, has to be adjusted to achieve the required $d_{A}(z_{0})$. The final model that would be wrongly inferred by matching the CMB acoustic scale, has $\O_{d}^{0}=0.735$ and is shown by the dashed line. {\emph{ Right:}} Fractional difference in the comoving angular diameter distance $d_{A}(z)$ from the  fiducial model. The dotted line represents the fractional error for the same model as shown by the dotted line on the left plot. As expected the difference in this case goes to zero at low redshift. The dashed line shows fractional difference in the wrongly inferred model. Note that since this model had its $\O_{d}^{0}$ shifted low, it overestimates the true distances by $\sim 1.5\%$ for $z\gtrsim 1$ and by $1.3\%$ for $z\lesssim 0.5$. }
\label{dAEDE}
\end{center}
\end{figure*}

Let us choose the fiducial dark energy model as one with $\O_{d}^{0}= 0.742$, $w_{0}=-1$ and $\ode = 0.03$ with the other cosmological parameters set at the values described in the body of the paper.  Now, if we had wrongly assumed  $\ode=0$ i.e. a $\Lambda$ model, then by fitting the CMB, we would find a set of parameters that would keep the angle $\theta_{A} = r_{s}(z_{0})/d_{A}(z_{0})$ constant. As $r_{s} \propto (1-\ode)^{1/2}$, we would find a model that overestimates the sound horizon and hence the comoving angular diameter distance to the last scattering surface by the factor $(1-0.03)^{-1/2} \simeq 1.015$ i.e. by $1.5\%$.  In particular, if we also impose the strong constraint that $\Om h^{2}$ is a constant, this model would have a value of  $\O_{d}^{0}$ that would be slightly lower than the true value, namely $\sim 0.735$. This would, in turn, make us overestimate the value of $d_{A}(z)$ to all redshifts (see Fig.~\ref{dAEDE}). \par
\begin{figure*}[htbp]
\begin{center}
\includegraphics[scale=0.39]{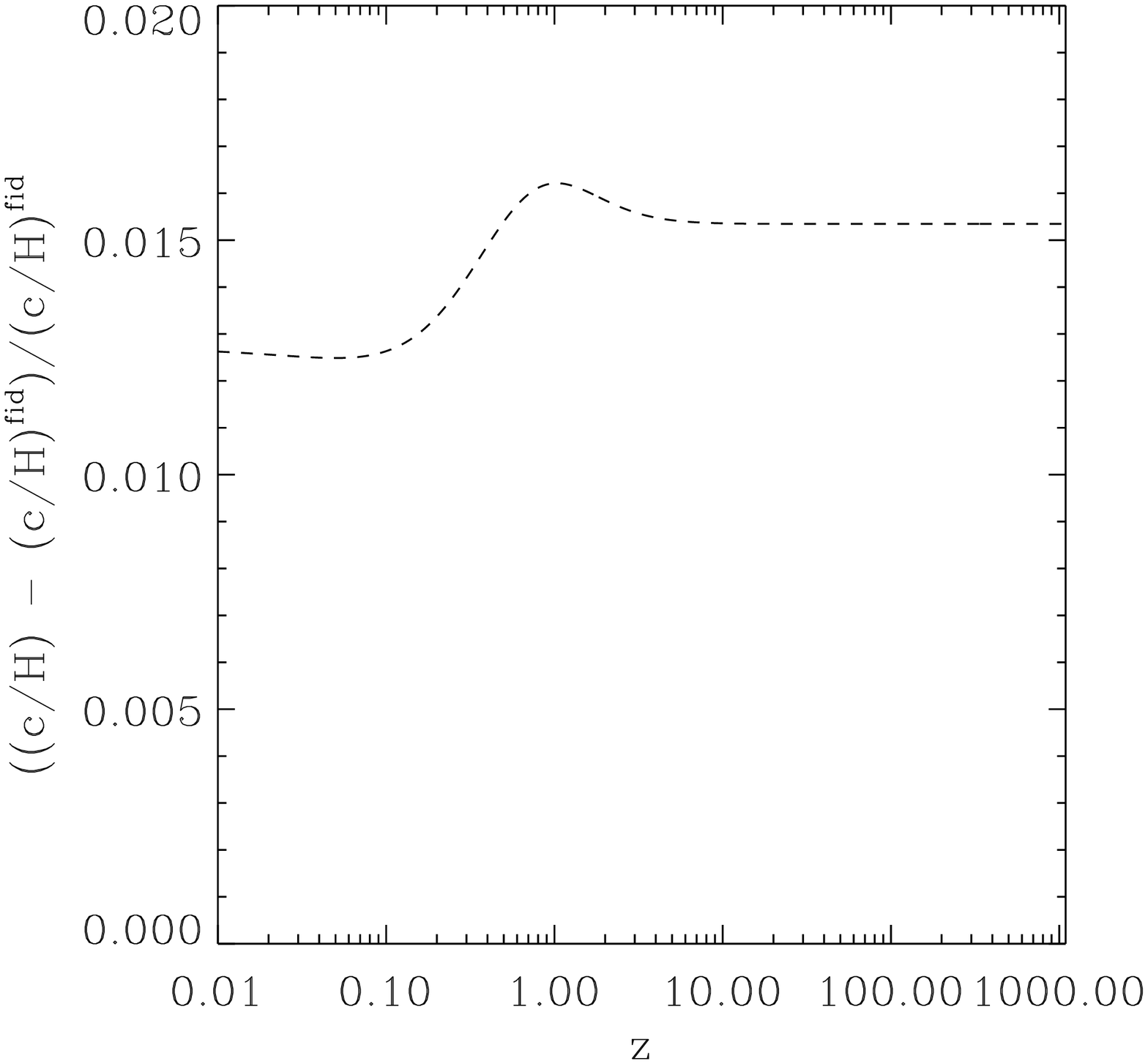}\includegraphics[scale=0.39]{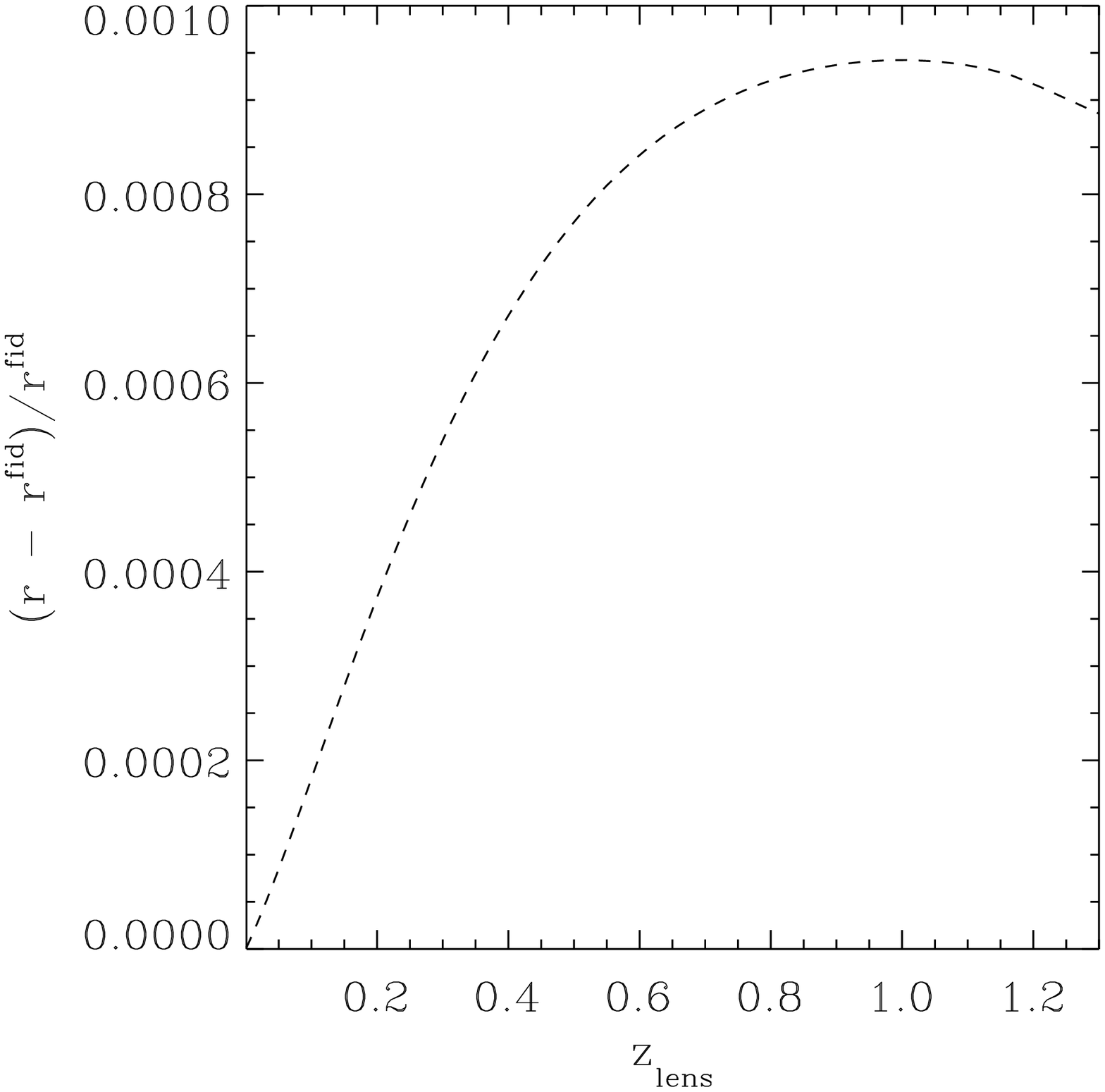}
\caption{{\emph{Left:}} Fractional difference from the fiducial of the  Hubble scale $c/H(z)$ in the wrongly inferred model of Fig.~\ref{dAEDE}. {\emph{Right:}} Same as left, for the lensing ratio as a function of the lens redshift.}
\label{deltaHubbleLens}
\end{center}
\end{figure*}

Now consider the BAO ratio at some redshift $z_{BAO}$. For the sound horizon at the drag epoch, we would again make an overestimate by the same factor, $\sim 1.5\%$. But at the same time, we  overestimate  $d_{A}(z_{BAO})$ or $c/H(z_{BAO})$ by an almost similar factor because $\O_{d}^{0}$ has shifted (see Figs.~\ref{dAEDE} \& Figs.~\ref{deltaHubbleLens}). Therefore, the transverse ratio  $r_{s}(z_{D})/d_{A}(z_{BAO})$ or the line-of-sight ratio $r_{s}(z_{D})H(z_{BAO})$ estimated in the wrong model will be rather close to the true values, thereby making it hard to detect the presence of early dark energy.  In fact, these figures show that deviations from the fiducial ratio occur only at the $0.2\%$ level at $z\lesssim 0.5$.  Because the angular scale of the CMB acoustic peak is a precisely measured quantity, it is expected that even if the strong CMB prior is relaxed, a MCMC type exploration of the parameter space would reveal a similar insensitivity of the BAO to $\ode$ \citep{2007JCAP...04..015D}.  
\par For the lensing ratio, the situation is even worse, as depicted in the right panel of  Fig.~\ref{deltaHubbleLens}. Here, due to overall shift of all distance scales, the difference between the inferred and the true model is less than $0.1\%$ at all lens redshifts. 

\end{document}